\documentclass[11pt]{article}

\usepackage[latin1]{inputenc}

\usepackage{amstext,amsmath,amssymb,amsfonts,bbm,amsthm}
\usepackage{hyperref}
\usepackage{authblk}
\usepackage{caption} 
\usepackage{epsfig} 
\usepackage{subfigure}
\usepackage{color} 
\usepackage{graphicx} 
\usepackage{braket} 
\usepackage{psfrag}  

\usepackage{float}  

\usepackage[lmargin=50pt,rmargin=60pt,tmargin=60pt,bmargin=65pt]{geometry}

\newcommand{\be}{\begin{equation}}
\newcommand{\ee}{\end{equation}} 
\newcommand{\Tr}{{\rm Tr}}
 
\DeclareMathOperator{\im}{\mathrm{i}}

\allowdisplaybreaks[4]

\newtheorem{theorem}{Theorem}

\theoremstyle{remark}

\begin{document}

\title{\bf On the generalization of the Wigner semicircle law to real symmetric tensors}
 
\author[1,2]{Razvan Gurau}
 
\affil[1]{\normalsize \it 
 CPHT, CNRS, Ecole Polytechnique, Institut Polytechnique de Paris, Route de Saclay, \authorcr 91128 PALAISEAU, 
 France, rgurau@cpht.polytechnique.fr
 \authorcr \hfill }

\affil[2]{\normalsize\it  
Perimeter Institute for Theoretical Physics, 31 Caroline St. N, N2L 2Y5, Waterloo, ON,
Canada
 \authorcr \hfill}

\date{}

\maketitle

\hrule\bigskip

\begin{abstract}
We propose a simple generalization of the matrix resolvent to a resolvent for real symmetric tensors 
$T\in \otimes^p \mathbb{R}^N$ of order $p\ge 3$. The tensor resolvent yields an integral representation for a class of tensor invariants and its singular locus can be understood in terms of the real eigenvalues of tensors.

We then consider a random Gaussian (real symmetric) tensor. We show that in the large $N$ limit the expected resolvent has a finite cut in the complex plane and that the associated ``spectral density'', that is the discontinuity at the cut, obeys a universal law which generalizes the Wigner semicircle law to arbitrary order. 

Finally, we consider a spiked tensor for $p\ge 3$, that is the sum of a fixed tensor $b\,v^{\otimes p}$ with $v\in \mathbb{R}^N$ (the signal) and a random Gaussian tensor $T$ (the noise).
We show that in the large $N$ limit the expected resolvent undergoes a sharp transition at some threshold value of the signal to noise ratio $b$ which we compute analytically.
\end{abstract}

\hrule\bigskip

%\newpage
\tableofcontents

\section{Introduction}

Recent progress in the study of large random tensors \cite{RTM,color,review} begs one to identify an appropriate notion of eigenvalues for a tensor and study their statistics. While no canonical notion of eigenvalues for tensors has so far emerged, several candidates have been proposed in the literature. In the terminology of Qi \cite{qi2005eigenvalues} we have $E$--eigenvalues, $Z$--eigenvalues, $H$--eigenvalues. Cui, Dai and Nie  discuss $D$--eigenvalues, or the more general ${\cal B}$--eigenvalues \cite{cui2014all}. Cartwright and Strumfels \cite{cartwright2013number} discuss yet another notion of eigenvalues (which they call simply eigenvalues) closely related to, but not identical with, Qi's $E$--eigenvalues\footnote{The $E$--eigenvalues of Qi \cite{qi2005eigenvalues} are the \emph{normalized eigenvalues} of Cartwright and Strumfels \cite{cartwright2013number}. However a tensor can have non normalized eigenvalues in the sense of \cite{cartwright2013number}}.
One thing is for sure: whichever notion of eigenvalues one prefers, computing all the eigenvalues of a tensor is an arduous task \cite{cui2014all}.
 
In this paper we define a resolvent function $\omega(w;T)$ for a real symmetric tensor $T\in \otimes^p \mathbb{R}^N$  of order
$p$\footnote{The number $p$ is often called the rank of $T$. As ``tensor rank''
is sometimes used in the literature to designate other aspects of a tensor, we prefer to call $p$ the order of $T$.}, that is a symmetric $p$--linear form over the vector space $\mathbb{R}^N$.
For $p=2$, $\omega(w,T)$ is the usual matrix resolvent
 $ N^{-1}\Tr [( w-T )^{-1} ]$ and has pole singularities in the complex plane of $w$ at the eigenvalues $\lambda_i$ of $T$. The spectral density is the empirical distribution of eigenvalues $\rho(y;T)\sim N^{-1}\sum_{i} \delta(y-\lambda_i) $. For larger $p$, $\omega(w;T)$ has  two infinite cuts from $0$ to $\pm \infty$. In this case the relation between the ``spectral density'', that is the discontinuity $\rho(y;T)$ of $\omega(w;T)$ at the real axis, and the (appropriate notion of) eigenvalues of $T$ is more involved. 
 In both cases the resolvent yields an integral representation for a class of invariants $I(T)$:
\be
    I (T)  = \int_\gamma \frac{dw}{2\pi \im} \; \omega(w,T) \; h_I(w) \; ,
\ee
where $\gamma$ is a simple contour which encircles the singular locus of $\omega(w,T)$ and $h_I(z)$ is a representation function associated to the invariant $I(T)$.
 
For all $p$, when considering the tensor to be random, distributed on a Gaussian\footnote{In recent work the largest eigenvalue of a real symmetric Gaussian random tensor has been studied \cite{Evnin:2020ddw}.}:
\be\label{eq:distribution}
 d\nu(T) \sim 
 \left[ \prod_{a_1 \le\dots \le a_p} d T_{a_1\dots a_p} \right]
   \exp\bigg\{
   -\frac{N^{p-1}}{2 p } \sum_{a_1\dots a_p=1}^N ( T_{a_1\dots a_p} ) ^2 \bigg\} \;,
\ee
and large, $N\to \infty$, the expectation of the resolvent displays a finite cut in the complex plane:
\be
\big[ -\sqrt{ p^{p}/(p-1)^{p-1}} , \sqrt{ p^{p}/(p-1)^{p-1}} \big] \;,
\ee
and the discontinuity at the cut obeys an universal law. For $p=2$ this is the Wigner semicircle law, and for $p\ge 3$ the generalized Wigner law can be written in terms of Meijer G--functions.
Due to the universality properties of the Gaussian measure  \cite{universality}, the same universal law is obtained for a large class of tensor probability measures. 

The surprising fact is that taking Gaussian expectations with respect to $T$ and going at large $N$ has a similar effect on the locus of singularities for any $p\ge 2$, although the starting point is very different at $p=2$. At $p=2$ the resolvent has poles and taking $T$ random (Gaussian) and large collapse the poles into a finite cut. For $p\ge 3$ the resolvent always has infinite cuts. Taking the tensor $T$ to be random (Gaussian) and large collapses the infinite cuts into a finite cut. 
 
 As an application, we study the resolvent in the case of a spiked tensor \cite{arous2019landscape} of order $p\ge 3$: 
\be
A = \frac{ b }{N^{\frac{p}{2}-1}} v^{\otimes p} + T \;,
\ee
 with $v$ a fixed vector in $\mathbb{R}^N$ of norm $1$ and $T$ a random Gaussian tensor as in Eq.~\eqref{eq:distribution}.
 We show that at the threshold value:
\be
 b_t^2 = \frac{(p-1)^p}{(p-2)^{p-2}} \; ,
\ee
of the signal to noise ratio the largest non removable singularity of the resolvent jumps from $\sqrt{ p^{p}/(p-1)^{p-1}}$ to $p^{p/2}$. 
We interpret this change in the singular locus of the resolvent as the detection of the spike and $b_t$ as the detection threshold. Our threshold value is to be compared to the one  obtained in Proposition 2 in \cite{arous2019landscape}.

A second more difficult question one asks is the reconstruction of the signal, i.e. once the presence of the spike is detected one should endeavour to find the fixed vector $v$. This goes beyond what we are able to do at the moment using the resolvent formalism discussed here, but we plan to address this question in the future. In the case $p=2$ the largest non removable singularity of the resolvent is the largest eigenvalue of $T$. One can study the statistics of the corresponding eigenvector and its correlations to the signal $v$.  It is a tantalising question whether a similar relation holds for higher $p$. This is in fact not impossible: although the singular locus of the resolvent for fixed $T$ is very different in the case $p \ge 3$, and in particular it exhibits cuts (hence there is no largest non removable singularity to associate to some eigenvalue), it is not impossible that after averaging and taking $N$ large, the newly created largest non removable singularity is in fact related to the largest eigenvalue of the tensor. One reason to hope that this is the case is that, as we explain below, in the Feynman expansion of the resolvent, the effect of taking $N$ large is to restrict to melonic graphs. If one restricts ad hoc to melonic graphs at fixed $T$, the amplitude of each graph should be dominated by the larges eigenvalue of $T$ and the radius of converge of this truncated resolvent should essentially be the largest eigenvalue. It is just that at finite $N$ this effect is drowned in the super exponential growth of number of Feynman graphs, which crushes the radius of convergence to zero. 
Studying this in detail is left for future work.

\section{Tensors, eigenvalues and invariants}

Let $p\ge 2$ be an integer and $T\in \otimes^p \mathbb{R}^N$ a real symmetric tensors of order $p$ transforming in the (symmetric) fundamental representation of the orthogonal group ${\cal O}(N)$. In components:
\be
 T_{a_1\dots a_p} \in \mathbb{R} \;,\quad
T_{b_1\dots b_p} = \sum_{a_1, \dots a_p=1}^N O_{b_1a_1} \dots 
 O_{b_p a_p} T_{a_1\dots a_p} \;, \qquad T_{a_1\dots a_p} = T_{a_{\sigma(1)}\dots a_{\sigma(p)}} \;,
\ee
for $O \in {\cal O}(N)$ an orthogonal transformation and $\sigma$ any permutation of $p$ elements.
For $x,y\in \mathbb{R}^N$ (or $\mathbb{C}^N$) we denote $x y \equiv \sum_{i=1}^N x_i y_i $ and for a (real symmetric) tensor $T$ and a vector $x\in \mathbb{R}^N$ (or $\mathbb{C}^N$) we denote:
\be
T x^p  \equiv \sum_{a_1, \dots a_p =1}^N T_{a_1\dots a_p } x_{a_1} \cdots x_{a_p}  \;, \quad 
 \big( Tx^{p-1} \big)_{a_1} \equiv \sum_{a_2,\dots a_p =1}^N T_{a_1 a_2 \dots a_p } x_{a_2} \cdots x_{a_p}  \;,
\ee
and $\im  = e^{\im \frac{\pi}{2}}$ the imaginary unit. 

\subsection{Eigenvalues for tensors}

For $p=2$, $T$ is a real orthogonal matrix and has $N$ real eigenvalues $\lambda_{i}$ (some of which might be equal) with $N$ associated linearly independent real eigenvectors $x_{i}$. Cartwright and Strumfels \cite{cartwright2013number},
expanding on work by Qi \cite{qi2005eigenvalues,qi2007eigenvalues}, call a pair $(\lambda \in \mathbb{C},x\in \mathbb{C}^N)$ a 
normalized eigenvalue and corresponding eigenvector of an order $p\ge 3$ tensor $T$ if:
\be\label{eq:eigen}
 Tx^{p-1} = \lambda x \;,\qquad x^2 = 1 \;. 
\ee
We stress that the eigenvalues and eigenvectors of a real symmetric tensor can be complex and that the normalization of the eigenvectors is $x^2=1$, not $x\bar x=1$ (where the bar denotes complex conjugate). Computing the eigenvalues of a tensor is a hard problem \cite{cui2014all}. 
For $p\ge 3$, if $(\lambda, x) $ is a normalized eigenpair then, due to the normalization condition, $e^{\im \alpha} x$ is also an eigenvector if and only if $e^{\im \alpha} =\pm 1$. The eigenvalue corresponding to $-x$ is $\lambda$ for $p$ even and $-\lambda$ for $p$ odd.

There are a number of subtle points about the eigenpairs of tensors. For instance a tensor can have a continuous infinity of complex eigenvectors corresponding to the same eigenvalue. Consider the following real symmetric tensor of order $3$ (discussed also in \cite{cartwright2013number}) and the equations determining its normalized eigenpairs:
\be
 Tx^3  =  2x_1^3 + 3x_1x_2^2 +3x_1x_3^2   
 \;,\qquad \begin{cases}
             & 2 x_1^2 + x_2^2 + x_3^2 = \lambda x_1 \;, \\
             & 2x_1 x_2 = \lambda x_2  \;, \\
             & 2x_1 x_3 = \lambda x_3 \;, 
           \end{cases} 
           \qquad 
           \text{and} \;\; x_1^2 + x_2^2 + x_3^3 = 1 \;. \\
\ee
This tensor has two eigenvalues $\lambda = \pm 2$, each with a continuous infinity of associated complex eigenvectors.  $( \pm 1, \im \alpha ,\alpha), \, \forall \alpha \in \mathbb{C}$. If in this example one uses the alternative normalization $x\bar x =1$, one gets only one normalized eigenvector for each eigenvalue but a continuous infinity of eigenvalues. We stick to the  Cartwright and Strumfels \cite{cartwright2013number} normalization $x^2=1$ because this ensures that a real symmetric tensor has
at most $[ (p-1)^N -1 ] / (p-2)$ normalized eigenvalues (Theorem 5.5 in \cite{cartwright2013number}). The bound is saturated for generic tensors. 

Following \cite{cartwright2013number}, we observe that $x\in \mathbb{C}^N$ with $x^2=1$ is an eigenvector of $T$ if and only if it is a critical point of the function $ Tx^p/p$ restricted to the affine hyper surface $x^2=1$. Indeed, recalling the method of Lagrange multipliers, a point is a critical point of a function subjected to a constraint only if the gradients of the function $\nabla ( Tx^p/p) = Tx^{p-1} $ and the constraint $\nabla(x^2-1) = 2x$ are proportional. This plus the constraint itself yield
Eq.~\eqref{eq:eigen}. The converse is trivial: any normalized eigenpair of $T$ is a critical point of $ Tx^p/p$ restricted to $x^2=1$. 

\paragraph{Real eigenpairs.}
If a normalized eigenvector of a tensor $T$ is real, $x\in \mathbb{R}^N$, then its corresponding eigenvalue is real $\lambda\in \mathbb{R}$. In this case we call 
$(\lambda,x)$  a real eigenpair\footnote{In \cite{qi2005eigenvalues} real eigenpairs are called $Z$-eigenpairs.}. Real symmetric tensors always have at least two real eigenpairs \cite{qi2007eigenvalues}. Indeed, a continuous function (that is $Tx^p$) on a compact set with no boundary (that is $x^2=1,\, x\in \mathbb{R}^N$) always attains its extrema and the extrema are critical points.
In fact they are expected to have many, as the real critical points of $Tx^p$ with $x^2=1$ yield a realization of the landscape of the spherical $p$--spin model \cite{arous2019landscape,Ros_2020}. For a random Gaussian tensor, with which we deal below, the expected number of real eigenpairs is exponential 
\cite{breiding2019many,breiding2017expected}.

\subsection{Invariants}

One can build many ${\cal O}(N)$ invariants starting from a tensor $T$. A particular class, the so called \emph{trace invariants} \cite{RTM} are, up to symmetries, in bijection with $p$--valent graphs. These are graphs such that every vertex has degree exactly $p$. Self loops and multiple edges are allowed. 

Let $b$ be a $p$--valent graph. To every vertex $v$ of $b$ we associate a tensor $T_{a^v_1 \dots a^v_p}$ and to every edge $e = (v,w)$ we associate the contraction of one index of $T_{a^v_1 \dots a^v_p}$ with one index 
of $T_{a^w_1 \dots a^w_p}$: 
\be\label{eq:trinv}
 \Tr_b(T) = \sum_a \prod_{v\in V(b) }T_{a^v_1 \dots a^v_p}
    \prod_{e= (v,w) \in E( b)}  \delta_{a^v_q a^w_{q'}} \;,
\ee
with $V(b)$ and $E(b)$ the vertex and edge sets of $b$. As the tensors we consider are symmetric, the choice of the indices $q$ and $q'$ is irrelevant, as long as for every vertex exactly one index is associated to every edge incident to it. The invariant is called connected, or single trace, if the graph $b$ is connected.

At finite $N$ the trace invariants are an over complete set. Any invariant can be decomposed as a linear combination of (possibly disconnected) trace invariants by averaging over the orthogonal group and using Weingarten calculus \cite{ColSni}. For non symmetric tensors, the suitably adapted trace invariants become an algebraic basis in the space of invariants at infinite $N$ \cite{BenGeloun:2017vwn}. We will assume the same holds for symmetric tensors.

In order to simplify the combinatorics it is convenient to work with 
combinatorial maps (or embedded graphs), that is graphs endowed with an ordering of the edges around the vertices. In detail, a \emph{combinatorial map} is:
\begin{itemize}
 \item[-] a finite set $H= \{ h_1, \dots h_{r}  \}$ of half edges,
 \item[-] a permutation $\tau$ on $H$ defining the successor half edge around the vertices. The cycles of $\tau$ are the vertices of the combinatorial map,
 \item[-] an involution $\tau'$ on $H$ with no fixed points defining the edges $(h,\tau'(h))$ of the map\footnote{Although they play no role below, we mention that the cycles of the permutation $ \tau \tau'$ are called the faces of the map.
}.
\end{itemize}

A combinatorial map is \emph{rooted} if one of its edges is marked by an arrow. The rooted tri--valent combinatorial maps with two vertices are depicted in figure \ref{fig:mapps} below.

\begin{figure}[htb]
\begin{center}
\includegraphics[width=0.5\textwidth]{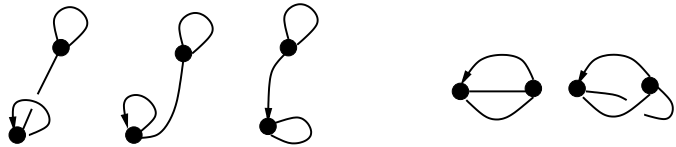}
 \caption{Connected rooted tri--valent combinatorial maps with two vertices.} \label{fig:mapps}
 \end{center}
\end{figure}

There are many combinatorial maps underlying the same graph, hence several rooted maps will correspond to the same trace invariant.
For instance the first three maps on the left figure \ref{fig:mapps}
correspond to the invariant $\sum_{a,b,c}T_{aab} T_{bcc}$ while the two rightmost ones correspond to $\sum_{a,b,c}T_{abc}T_{abc}$.

\section{The Gaussian $p$--spin model}

We denote $ [d\phi] =(2\pi)^{-N/2}  \prod_{i=1}^N  d\phi_i$. 
The Gaussian $p$--spin model is defined by the partition function:
\be\label{eq:vec}
 {\cal Z}(w;T) =  \int_{\mathbb{R}^N} [d\phi ] \; e^{-{\cal S}(\phi)} \;,\qquad
 {\cal S}(\phi) =
 \frac{ \phi^2 }{2} - \frac{1}{w} \, \frac{T\phi^p}{p}  \;,
\ee
with $\phi$ a vector in $\mathbb{R}^N$, $T$ a real symmetric tensor of order 
$p$ and $w=|w| e^{\im \psi}$ with $-\pi<\psi<\pi$ a coupling constant.
Related models include the spherical $p$--spin model \cite{Berlin:1952zz,Stanley:1968gx} in which the spins are constraint to have norm one or the SYK model \cite{Sachdev:2011cs,Kitaev} with fermions. 
The perturbative expansion of ${\cal Z}( w;T)$ around $w = \infty$  is obtained by Taylor expanding the interaction:
\be
 {\cal Z}(w;T) = \sum_{n\ge 0} \frac{1}{n!} \, \frac{1}{w^n} \, \int_{\mathbb{R}^N} [d\phi] \; 
e^{-\frac{1}{2} \phi^2} \left( \frac{T\phi^p}{p} \right)^n \;.
\ee
It is a standard fact \cite{Gurau:2017xhf,Bonzom:2018jfo} that performing the Gaussian integral one obtains an expansion of 
${\cal Z}(w;T)$ in terms of (possibly disconnected) combinatorial maps corresponding to the trace invariants:
\be\label{eq:seriestosum}
 {\cal Z}(w;T) =  \sum_n \frac{1}{n!} \sum_{ b \in {\cal M}_n} 
     \frac{1}{w^n} \; \Tr_b(T) \;,
\ee
where ${\cal M}_n$ denotes the set of maps with $n$ vertices labelled $1,\dots n$. There are $   (np)! / (\frac{np}{2})!    $
such maps.

In order to study the analyticity properties of ${\cal Z}(w;T)$ as a function of the coupling constant $w$ we mimic\footnote{The only subtlety is that in order to bound the Taylor rest term for the $p$--spin model we use the Cauchy Schwartz inequality $
 | T\phi^p | \le \sqrt{  (TT)} \sqrt{ (\phi^p \phi^p) }
  = \sqrt{T^2} \; (\phi^2)^{p/2}
$. Tracking the factors of $N$ we obtain (see Appendix \ref{sec:toymodel} for the notation) a bound:
\be
\big|R_q^{(n)}(g) \big| \le \frac{1}{n!} \;\frac{ |g|^{n\frac{p-2}{2} } }{p^n} \; (T^2)^{n/2} \; N^{\frac{np}{2}} \frac{(np)!}{ 2^{\frac{np}{2} } (np/2)!  }  \;  \frac{1}{ 
  \bigg( \cos\bigg[ \frac{p-2}{p} (\alpha_q-\alpha) \bigg] \bigg)^{np+1/2}  }  \; .
\ee
This suffices for proving Borel summability at finite $N$.} the steps in Appendix \ref{sec:toymodel}. The partition function is best seen as the directional Borel Leroy sum of its perturbative series. It is analytic outside the real axis and can represented in the upper (respectively lower) complex 
half plane as:
\be
\begin{split}
0 <\psi<  \pi \;\qquad & {\cal Z}_+(w;T)= 
 \int_{e^{\im \theta_+}\mathbb{R}^N} [d\phi] \;\exp \bigg\{ - \left( \frac{\phi^2}{2} - \frac{1}{w} \, \frac{T\phi^p}{p}\right) \bigg\} \;, 
  \qquad \theta_+ = \frac{1}{p} \left( \psi - \frac{\pi}{2}\right) \;,
 \\
-\pi < \psi< 0  \;\qquad & {\cal Z}_-(w;T)=
\int_{e^{\im \theta_-}\mathbb{R}^N} [d\phi] \;\exp \bigg\{ - \left( \frac{\phi^2}{2} - \frac{1}{w} \, \frac{T\phi^p}{p}\right) \bigg\} \; ,
 \qquad \theta_- = \frac{1}{p} \left( \psi + \frac{\pi}{2}\right) \; .
\end{split}
\ee

The analytic structure of ${\cal Z}(w;T)$ is different for $p=2$. For $p\ge 3$ the partition function ${\cal Z}(w;T)$ has two\footnote{Unless $p$ is even and $T$ is positive semi defined (that is $T\phi^p \ge 0,\;\forall \phi$) or negative semi defined, in which case only one of the cuts is present as all the eigenvalues of real eigenpairs have the same sign \cite{qi2005eigenvalues}.} cut singularities along the positive and the negative real axis. Its discontinuity at $y\in \mathbb{R}$ is:
\be\label{eq:discinst1}
     {\cal Z}_{+}( y ;T ) -  {\cal Z}_{ - }( y ;T ) 
      =\sum_{\nu} \frac{1}{\sqrt{ {\cal S}''(\phi_{\nu} ) }} e^{-{\cal S}(\phi_{\nu} ) } \;,
\ee
where $\phi_{\nu}\in \mathbb{R}^N$ are the real instantons, that is the solutions of the equation of motion ${\cal S}'(\phi_{\nu})=0$ which are real for $w =y \in \mathbb{R}$.
The ambiguity in the determination of the square root in Eq.~\eqref{eq:discinst1} is fixed by checking carefully the steepest descent paths through the instantons
(see Appendix \ref{sec:toymodel} for a detailed example).

As $\phi_{\nu}\in \mathbb{R}^N$ and $\phi_{\nu} \neq 0$ we have that $\phi_{\nu}^2$ is real and strictly positive. It follows that $\phi_{\nu}$ is a real instanton at $y$ if and only if there exists a real eigenpair 
$(\lambda_{\nu},x_{\nu})$ of $T$ such that:
\be
  x_{\nu} =  \frac{\phi_{\nu}}{(\phi_{\nu}^2)^{1/2}} \;, \qquad 
  \lambda_{\nu} = \frac{y}{ (\phi_{\nu}^2)^{1/2} } \;.
\ee

We observe that for $y\in \mathbb{R}_{+}$ the real instantons $\phi_{\nu}$ are in one to one correspondence with the real eigenpairs $(\lambda_{\nu}, x_{\nu})$ with positive eigenvalue $\lambda_{\nu} > 0$, while  for $y\in \mathbb{R}_{-}$ the real instantons $\phi_{\nu}$ are in one to one correspondence with the real eigenpairs with negative eigenvalue $\lambda_{\nu} < 0$. 

For odd $p$, if $(\lambda,x)$ is an eigenpair then $(-\lambda ,-x)$ is an eigenpair. Denoting $\lambda_{\max} $ the largest eigenvalue of a real eigenpair of  $T$, which is necessarily positive, we get: 
\be
{\cal Z}_+( y ) -  {\cal Z}_- ( y  )  
\sim e^{-\frac{p-2}{2p} \left( \frac{ |y| }{ \lambda_{\max} }\right)^{\frac{ 2}{p-2} } } + \dots \;.
\ee

For $p$ even denoting $\lambda_{\max} >0$ and $\lambda_{\min}<0$  the largest and smallest eigenvalues of the real eigenpairs of $T$ we have:
\be 
{\cal Z}_+( y ) -  {\cal Z}_-( y  ) 
 \sim  \begin{cases}
    e^{-\frac{p-2}{2p} \left( \frac{ y }{ \lambda_{\max} }\right)^{\frac{ 2}{p-2} } } + \dots \;,\qquad & y>0  \\
    e^{-\frac{p-2}{2p} \left( \frac{ y }{ \lambda_{\min} }\right)^{\frac{ 2}{p-2} } } + \dots \;,\qquad & y<0
   \end{cases} \;.
\ee

We have so far discussed the analytic structure of the partition function.
The main point is that  (see section 3.5 in \cite{Marino:2015yie})
all the correlation functions have the same analytic structure in $w$: they are analytic outside the real axis and exhibit two  infinite cuts from $0$ to $\pm \infty$.

\subsection{The resolvent} 
 
We now define a resolvent function for a tensor $T$ of order $p\ge 2$. 
The crucial point is that the resolvent we introduce below yields a spectral representation for a class of tensor invariants.

\paragraph{The resolvent for matrices.} 
Let us recall the resolvent of matrices. A symmetric $N\times N$ matrix $T$ has 
$N$ real eigenvalues $\lambda_{i}\in \mathbb{R}$. The resolvent of $T$ is 
the matrix $(w-T)^{-1}$ and is well defined as long as $w\neq \lambda_{i}$\footnote{In infinite dimension the resolvent becomes an operator
and is well defined for $w$ in the resolvent set, the complement of the spectrum of $T$.}. The resolvent defines projectors on the eigenspaces of $T$, that is for any function $F$ we have:
\be
 F(T) = \int_{\cup_i \gamma_{i} } \frac{dw}{2\pi\im}\; 
    F(w) \, \frac{1}{w-T} \;, 
\ee
where the contour of integration is the union of positively oriented circles $\gamma_{i}$ centered at the eigenvalues $\lambda_i$. If $F$ is holomorphic we can join the circles into a unique contour $\gamma$ encircling all the eigenvalues. 
The normalized trace of the resolvent matrix is the resolvent function:
\be
\omega(w;T) = \frac{1}{N} \, \Tr\left( \frac{1}{w-T} \right) \; .
\ee
This is a meromorphic function of $w$ with simple poles at $\lambda_i$ and with residues the multiplicities. A spectral representation of the single trace invariants of $T$ (traces of functions of $T$) is:
\be
 \frac{1}{N} \, \Tr\big[ F(T)\big] = \int_{\gamma}  \frac{dw}{2\pi\im}\; 
    F(w)  \, \omega(w;T) \; = 
    \sum_{n\ge 0 }\int_{\gamma} \frac{dw}{2\pi\im}\; 
    \frac{ F(w) }{w^{n+1}} \, \frac{  \Tr(T^n) }{N} \;   \; ,
\ee
where $\gamma$ encircles all the eigenvalues of $T$ and the origin.

\paragraph{A resolvent for tensors.}

We define the \emph{balanced resolvent} of the tensor $T$ 
as the two point function of the Gaussian $p$--spin model, that is the connected expectation of $\phi_a \phi_b$ with the measure in Eq.\eqref{eq:vec}. Its normalized trace is:
 \be\label{eq:defres}
 \omega(w;T) = 
  \frac{w^{-1}}{ {\cal Z}(w;T)  } \int_{\mathbb{R}^N} [d\phi] \; 
   \frac{\phi^2}{N}  \;
  \exp \bigg\{ - \bigg(\frac{1}{2} \phi^2  - \frac{ 1 }{w} \; \frac{ T \phi^p }{p}  \bigg) \bigg\} \; .
\ee

This is well defined outside the real axis. For $p\ge 3$ it has two cuts from $0$ to $\pm\infty$. The balanced resolvent function $\omega(w;T)$ respects:
\be\label{eq:SDE}
 0= \frac{w^{-1}}{ {\cal Z}(w;T)  }   \int [d\phi] \sum_i  
 \frac{1}{N} \, \frac{\delta}{\delta \phi_i}
    \bigg[ \phi_i e^{  - \big(\frac{1}{2} \phi^2  - 
    \frac{ 1}{w} \; \frac{ T \phi^p }{p} \big) } \bigg]
    = w^{-1} - \omega(w;T) - \frac{p}{N} \frac{d}{dw} 
    \ln{\cal Z}(w;T) \;.
\ee
We note that for $p=2$ our definition yields by direct integration:
\be
   \omega(w;T)  =\frac{1}{w} + \frac{1}{N} \frac{d}{dw} 
   \Tr \ln\left(1-\frac{T}{w}\right)
    = \frac{1}{N} \Tr  \left( \frac{1}{w-T} \right) \;,
\ee
that is the balanced resolvent function reduces to the usual resolvent for matrices.

\paragraph{Invariants and the resolvent.}
We denote ${\cal B}_n$ the set of connected rooted $p$--valent maps with $n$ unlabelled vertices and we define the \emph{balanced} single trace invariant of degree $n$ as:
\be
 I_n(T) =    \sum_{b\in {\cal B}_n } \Tr_b(T) \;,
\ee
where we emphasize that each rooted map is counted with weight 1. For $p=2$ there is only one connected rooted two--valent map with $n$ vertices: the rooted cycle of length $n$. Thus for a symmetric matrix $T$ we have 
$I_n(T) = \Tr(T^n)$.
For higher order tensors the cardinal of ${\cal B}_n$ is of order 
$n^{ n \frac{p-2}{2}}$. For instance, taking into account Fig.~\ref{fig:mapps}, we have for a tensor of order $3$:
\be
I_2(T) = 3\sum_{a,b,c} T_{aab} T_{bcc} + 2 \sum_{a,b,c}T_{abc} T_{abc} \;.
\ee
An arbitrary balanced invariant is an element in the span of the balanced invariants of arbitrary degree (properly normalized with $N$):
\be
 I(T) = \sum_n a_n  \; \frac{ I_n(T) }{N}   \; .
\ee
For $p=2$ all the single trace invariants are balanced, but for $p\ge 3$ only a subclass is, and the space of invariants is much richer.
To each invariant $I(T)$ we associate a function:
\be\label{eq:f_I}
h_I(w) = \sum_n a_n w^n \;,
\ee 
which is analytic in some suitable domain in the complex plane.

The resolvent function defined in Eq.~\eqref{eq:defres} is the generating function of $ I_n(T) /N $. To see this, recall the Feynman expansion of the partition function ${\cal Z}(w;T)$ is an expansion in combinatorial maps. Taking the logarithm leads to a sum over connected maps, the derivative with respect to $z$ marks a vertex, and the factor $p$ chooses a half edge on the vertex, thus from Eq.~\eqref{eq:SDE} we get:
\be\label{eq:ressum}
 \omega(w;T) = \frac{1}{w} - \frac{p}{N}
 \frac{d}{dw} \ln {\cal Z}(w;T) = \sum_{n\ge 0} \frac{1}{w^{n+1}}
   \;\frac{1}{N} \sum_{b\in {\cal B}_n } \Tr_b(T) =  \sum_{n\ge 0} \frac{1}{w^{n+1}}
   \;\frac{  I_n(T) }{N} \;,
\ee
where by convention $I_0(T) = N$. Consequently any balanced invariant supports the spectral representation:
\be\label{eq:spectrbal}
  I(T) = \int_{\gamma } \frac{dw}{2\pi\im}\; 
    \omega(w;T) \;  h_I(w)   \; ,
\ee
where $\gamma$ encircles the real axis, that is $\gamma = \gamma_+ \cup \gamma_-$ with $\gamma_+$ going from $\infty$ to $-\infty$ in the upper complex half plane and 
$\gamma_-$ going from $-\infty$ to $+ \infty$ in the lower complex half plane.

The subtle point here is that the Feynman expansion is not convergent for $p\ge 3$. Indeed, as the cardinal of ${\cal B}_n$ behaves like $n^{n\frac{p-2}{2}}$, the series in Eq.~\eqref{eq:ressum} has zero radius of convergence, and $\omega(w,T)$ must be defined by directional Borel Leroy summations.

\section{The tensor Gaussian Orthogonal Ensemble}

We now consider $T$ to be a tensor drawn from the Gaussian ensemble:
\be\label{eq:pGOE}
 d\nu(T) = Q_N 
 \left[ \prod_{a_1 \le\dots \le a_p} d T_{a_1\dots a_p} \right]
   \exp\bigg\{
   -\frac{N^{p-1}}{2 p } \sum_{a_1\dots a_p=1}^N ( T_{a_1\dots a_p} ) ^2 \bigg\} \;,
\ee
where $Q_N$ is chosen such that the measure is normalized.
This is the straightforward generalization of the 
Gaussian Orthogonal Ensemble (GOE) to tensors of order $p$. The GOE is recovered for $p=2$ and is our benchmark.
We denote by $\Braket{\dots}$ the expectation with respect to $d\nu(T)$:
\be
 \Braket{g(T) }  = \int d\nu(T) \; g(T) \; .
\ee

We aim to study the statistics of $\omega(w;T)$ and of its singular locus in the large $N$ limit for $T$ drawn form the Gaussian ensemble in Eq.~\eqref{eq:pGOE} as
this captures the statistics of arbitrary balanced trace invariants
via the spectral representation in Eq.~\eqref{eq:spectrbal}.
Let us denote the expectation of the resolvent by:
\be
 \omega_{\pm}(w)  = \Braket{ \omega_{\pm}(w;T) } =
 \frac{1}{w}  -\frac{ p }{N}  \frac{d}{dw}
  \Braket{ \ln {\cal Z}_{\pm}(w;T) } \;,
\ee
which involves the derivative of the quenched (with respect to the random tensor $T$) free energy of the vector model in Eq.~\eqref{eq:vec}. It turns out that at leading order in 
the large $N$ limit the quenched and annealed averages are equal\footnote{One can compute the quenched average using the replica trick $\ln{\cal Z} = \lim_{n\to 0} ( {\cal Z}^n-1 ) /n $ and note that the replica symmetry is not broken at leading order in $1/N$ \cite{lesieur2017statistical}. Alternatively one can use the Feynman expansion and prove that at large $N$ the melonic graphs \cite{critical} of the quenched model dominate also the annealed one \cite{Gurau:2017xhf,Bonzom:2018jfo}.}
hence at leading order:
\be
 \omega_{\pm}(w)  = \frac{1}{w}  -\frac{p}{N}  \frac{d}{dw}
 \ln  \Braket{  {\cal Z}_{\pm}(w,T) } \;.
\ee
It remains to compute the annealed free energy. For this we note that 
the Gaussian expectation of the partition function can be computed by a translation:
\be
     \Braket{ e^{ \frac{1}{w} \, \frac{ T\phi^p }{p} } } 
     = e^{  \frac{1}{ N^{p-1}} \,\frac{ (\phi^2)^p }{2p w^2} } 
     \Rightarrow  \Braket{  {\cal Z}_{\pm}(w;T) } =  
     \int_{e^{\im  \theta_{\pm}}\mathbb{R}^N } [d\phi] \; 
      \exp\bigg\{-\frac{1}{2} \phi^2 + \frac{1}{ N^{p-1} } \,\frac{ (\phi^2)^p}{2 p w^2} \bigg\}  \; ,
\ee
where we recall that $w = |w|e^{\im \psi}$ and $\theta_{\pm} = (\psi \mp \pi/2)/p$, hence the integral is convergent.
Changing variables to spherical coordinates and rescaling the radial coordinate $\phi^2= N \rho^2 $ yields:
\be\label{eq:intform}
\begin{split}
& \Braket{  {\cal Z}_{\pm}(w;T) }  = K_N
 \int_{e^{\im \theta_{\pm}}\mathbb{R}_+}^{\infty} d\rho \; \rho^{-1} 
   e^{N f(\rho)} \;, \crcr
& \qquad   f(\rho) = \ln\rho -\frac{1}{2} \rho^2 + \frac{1}{2p w^2 } \rho^{2p} \;,
   \;\; f'(\rho) = \frac{ 1 -  \rho^2  + \frac{1}{w^2}\rho^{2p} }{\rho}\;,
\end{split}
\ee
with $K_N$ some constant independent of $w$. In this form the annealed partition function can be computed by a saddle point approximation. The saddle point equation 
$\partial_{\rho} f(\rho) = 0$ admits a unique physical
solution (which remains at finite distance from the origin when sending $w\to \infty$):
\be
\rho_0^2 =    T_p( w^{-2}) \;,
\ee
where $T_p(u)$ is the Fuss-Catalan function which we now recall.

\paragraph{The Fuss-Catalan function.} The Fuss-Catalan function $T_p(u)$ is well known in the literature. Here we follow the presentation in \cite{Krajewski:2017thd}. $T_p(u)$ is defined by the implicit equation:
\be
T_p(u)  = 1  + u   T_p(u)^p\; , \qquad u(T) = T^{-p + 1} -T^{-p} \;,
\ee
supplemented by the condition that $T_p(u)$ is analytic at $u=0$.
By the implicit function theorem the equation can be solved up to the critical point $u_c = (p-1)^{p-1}/p^p$ (corresponding to $T_c = p/(p-1)$) where $\partial u/\partial T = 0$. 
For $|u|<u_c$ it admits the absolutely convergent series representation:
 \be
   T_p(u) = \sum_{n\ge 0 }  F_p(n) u^n \;, \qquad   
   F_p(n) = \frac{1}{pn + 1} \binom{pn+1}{n}
   \;,
 \ee
 where $F_p(n)$ are known as the Fuss--Catalan numbers. Moreover $T_p$ can be analytically continued to the cut complex plane $\mathbb{C} \setminus [u_c,+\infty)$ by noting  
 \cite{Krajewski:2017thd,penson2011product} that the Fuss--Catalan numbers admit the integral representation:
 \be
  F_p(n) = \int_0^{1/u_c} dx\; x^{n} P_p(x) \;,
 \ee
with $P_p(x)$ a real positive function. Formally $P_p(x)$ is 
the inverse Mellin transform of $F_p(\sigma)$, the analytic continuation of the Catalan numbers. It can be expressed in terms of the Meijer $G$--function \cite{penson2011product} or written explicitly in terms of hypergeometric functions (note that $p-1=s$ in the notation of \cite{penson2011product}):
\be\label{eq:Pp}
\begin{split}
   P_p(x)  = & \sum_{k=1}^{p-1} \Lambda_{k,p} \;
     x^{\frac{k -p }{p}}  {}_{p-1}F_{p-2}
     \bigg(  \bigg\{ 1 - \frac{1+j}{p-1} +\frac{k}{p}\bigg\}_{j=1}^{p-1}  ,  
     \bigg\{ 1 + \frac{k-j}{ p }\bigg\}_{\genfrac{}{}{0pt}{}{j=1}{ j\neq k } }^{p-1}   ; \frac{(p-1)^{p-1}}{p^p}x\bigg) \;, \crcr
 \Lambda_{k,p} = & \frac{1}{(p-1)^{3/2}} \sqrt{\frac{p}{2\pi}}
 \left( \frac{ (p-1)^{p-1 }}{p^p}\right)^{\frac{ k}{p} }
 \;   \frac{ 
   \prod_{j=1,\dots p-1}^{j\neq k } \Gamma\left( \frac{j-k}{p} \right) }
  { \prod_{j=1}^{p-1} \Gamma \left( \frac{j+1}{p-1} -\frac{k}{p}  \right) } \;.
\end{split}
\ee
In particular:
\be\label{eq:lowored}
 P_2(x) = \frac{\sqrt{ 1 - \frac{1}{2^2}  x }}{\pi \sqrt{x}} \;, \qquad
 P_3 (x) =  \frac{ 1 }{ 2 \pi \;  x^{2/3}  }  
  \left(  \frac{ 3^{1/2} }{ 2^{1/3} } \right)
 \frac{ \left( 1 +  \sqrt{ 1 -  \frac{ \cdot 2^2 }{3^3} x } \right)^{2/3} - \left( \frac{2^2}{3^3} x\right)^{1/3}}{
 \left(1 + \sqrt{1 - \frac{ \cdot 2^2}{3^3} x } \right)^{1/3} } \;.
\ee
The Fuss Catalan function admits for $u \in \mathbb{C} \setminus [u_c,\infty)$ the convergent integral representation:
\be
  T_p(u)  = \int_0^{1/ u_c } dx\; \frac{ 1 } { 1 - u x } 
  P_{p}(x) \; .
\ee

\subsection{The Wigner law for arbitrary order tensors}

The expectation of the resolvent is obtained by the saddle point method as:
\be
\omega(w) = w^{-1 } - p  \frac{d}{dw} f(\rho_0) = 
w^{-1} -p 
 \bigg[ \partial_w  f(\rho_0) +  \partial_{\rho} f ( \rho_0)  \frac{\partial \rho_0 }{\partial w}\bigg] = \frac{1}{w}\rho_0^2 
 = \frac{1}{w} T_p(w^{-2})
 \;,
\ee
where we used the saddle point equation. At low $p$ we get, denoting $w_c^2 = p^p/(p-1)^{p-1}$ (we select the root which behaves like $w^{-1}$ at large $w$):
\begin{align}
 p = 2: \quad & \omega(w) =  
  \frac{w}{2} \left( 1 -  \sqrt{1 - \frac{ w_c^2 }{w^2}} \right)
   \;,
 \crcr
 p = 3: \quad & \omega(w) 
 = \frac{ \im }{3^{1/2}}
 \bigg[   \left( \sqrt{ 1 - \frac{  w_c^2 }{w^2}} - \im \frac{ w_c }{w}  \right)^{1/3}  - 
\left( \sqrt{ 1 - \frac{ w_c^2}{w^2}  } + \im \frac{ w_c}{w} \right)^{1/3}   \bigg]  \; .
\end{align}
The function $\omega(w)$ has a cut $[-\sqrt{w_c}, \sqrt{w_c}]$ and we obtain the spectral density (discontinuity at the cut) using the Sokhotski--Plemelj formula $ \rho( y ) =( 2\pi \im )^{-1}
 \lim_{\epsilon\to 0} \big[\omega( y - \im \epsilon ) -
\omega( y + \im \epsilon )  \big]$:
\be\label{eq:loworders}
\begin{split}
p=2 \qquad  \rho(y) = & \frac{1}{2\pi} \sqrt{4-y^2} \;, \crcr
p=3  \qquad  \rho(y) = & \frac{1}{2\pi |y|^{1/3}}
 \left( \frac{3^3}{2^2} \right)^{1/6}
  \bigg[ 
  \bigg( 1 + \sqrt{ 1-\frac{y^2}{ 3^3/2^2 } } \bigg)^{1/3}
  - \bigg( 1 - \sqrt{ 1-\frac{y^2}{ 3^3/2^2 } } \bigg)^{1/3} 
  \bigg] \; ,
\end{split}
\ee    
where the first line reproduces the Wigner semicircle law.
For arbitrary $p$ we get the spectral representation:
\be
 \omega(w) = \frac{1}{w} \int_0^{w_c} dx\; \frac{1}{1-\frac{x}{w^2}} P_p(x) = \int_{ -\sqrt{w_c}}^{ \sqrt{w_c}} dy
 \;\frac{ |y| P_p(y^2) }{ w-y } \;,
\ee
therefore the expectation of the resolvent always has a cut located at $(-\sqrt{w_c} , \sqrt{w_c})$ and its the spectral density obeys the generalized Wigner law: 
\be
 \rho(y) = |y| P_p(y^2) \;,\qquad
    y \in \left( -\frac{p^{p/2}}{ (p-1)^{(p-1)/2} }  , \frac{p^{p/2}}{ (p-1)^{(p-1)/2} } \right) \; ,
\ee
with  $P_p$ defined in Eq.~\eqref{eq:Pp}. The particular cases $p=2,3$ in Eq. \eqref{eq:loworders} follow from this general formula taking into account Eq.~\eqref{eq:lowored}. 

\paragraph{Feynman expansion.} The expectation of the resolvent can be derived directly from the Feynman expansion in Eq.~\eqref{eq:ressum}:
\be
 \Braket{ \omega(w;T)  } = \sum_{n\ge 0} \frac{1}{w^{n+1}}
   \;\Braket{\frac{  I_n(T) }{N} }\;, \qquad
    I_n(T)  =    \sum_{b\in {\cal B}_n } \Tr_b(T)  \;.
\ee
The crucial remark is that in the large $N$ limit only the melonic invariants $b$ survive, and each such invariant has expectation $1$ \cite{universality}. As the number of rooted melonic graphs is given by the Fuss-Catalan numbers \cite{critical}, we directly obtain $ \omega(w) = T_p(w^{-2})$. 

The point is that this derivation is only formal, as the Feynman expansion diverges. As it turns out, taking the large $N$ limit in each term in the divergent Feynman expansion one gets a convergent series which reproduces the right leading order behavior.

\subsection{The spiked tensor model}

Throughout this section we set $p\ge 3$.
The spiked tensor model (see \cite{arous2019landscape,lesieur2017statistical} and references therein) consists in considering a tensor:
\be
 A_{a_1\dots a_p} =  \frac{ b }{N^{\frac{p}{2}-1}} \, v_{a_1} \dots v_{a_p} + T_{a_1\dots a_p} \;,
\ee
with $v$ a fixed vector in $\mathbb{R}^N$ with $v^2=1$ and $T$ distributed on the Gaussian ensemble in Eq.~\eqref{eq:pGOE}. The normalization of the spike stems from having a fixed spike $v$ rather than a prior probability distribution for $v$. The coefficient $b$ is the signal to noise ratio. A typical question in this context is for what values of $b$ is the spike 
detectable. 

We will compute the expectation of the resolvent of $A$. This is still given by the equation:
\be
\omega_{\pm}(w) = w^{-1} - pN^{-1} \frac{d}{dw}
\Braket{\ln{\cal Z}_{\pm}(w;A)} 
\;,
\ee
and quenched is still equal to annealed in the large $N$ limit \cite{lesieur2017statistical}. The annealed partition function is:
\be
  \Braket{  {\cal Z}_{\pm}(w;A ) } =  
     \int_{e^{\im \theta_{\pm}} \mathbb{R}^N } [d\phi] \; 
      \exp \bigg\{
       -\frac{1}{2} \phi^2 + \frac{1}{N^{p/2-1}} \frac{b}{pw}   (v\phi )^p +\frac{1}{ N^{p-1} } \,\frac{ (\phi^2)^p}{2 p w^2} 
      \bigg\} \;.
\ee
Using spherical coordinates with $v$ aligned along the $\phi_1$ 
axis\footnote{Recall that in spherical coordinates we have:
\[
 \phi_1 = \rho \cos \theta_1 \;, \;\; \phi_{2} = \rho \sin\theta_1 \cos\theta_2 \;,\dots \;,
 \phi_{N-1}=\rho \sin\theta_1 \dots \sin\theta_{n-2} \cos\theta_{N-1} \;,\;\; \phi_{N}= \rho \sin\theta_1 \dots \sin \theta_{N-1}  \;,
\]
and the volume is:
\[
 \int_0^{\infty} d\rho \int_0^{2\pi} d\theta_{N-1}\int_{0}^{\pi} d\theta_i \; 
  \rho^{N-1} \sin^{N-2}\theta_1  \dots   \sin^2\theta_{N-3} \sin\theta_{N-2}
   \;.
\]
}, that is $v\phi = \rho \cos \theta_1$, and rescaling the radial coordinate $\rho$ by $N^{1/2}$ we get:
\be 
\begin{split}
\Braket{  {\cal Z}_{\pm}(w;A) } = &  K'_N  \int_0^{\pi} d\theta  \int_{e^{\im \theta_{\pm} } \mathbb{R}_+ } d\rho \;  \rho^{-1} \sin^{-2}\theta \\
   & \qquad \exp\bigg\{ N \bigg( \ln (\sin \theta) + \ln \rho - \frac{1}{2}\rho^2 
  + \frac{b}{ p w} \, \rho^p \cos^p \theta + \frac{1}{2pw^2 } \rho^{2p} \bigg) \bigg\} \; ,
\end{split}
\ee
with $K'_N$ some constant. A saddle point approximation gives 
$\Braket{  {\cal Z}_{\pm}(w;A) }= Nf(\theta_{\star},\rho_{\star})$ with  $ (\theta_\star,\rho_\star^2)$ the dominant saddle point of:
\be
 f(\theta, \rho) = \ln ( \sin \theta) + \ln \rho -\frac{1}{2}\rho^2 + \frac{b}{wp} \, \rho^p \cos^p\theta  + \frac{1}{2pw^2 } \rho^{2p} \;.
\ee

Unsurprisingly, for $b=0$ the $\theta$ integral decouples and we fall back in the previous case. For general $b$ we compute the  derrivatives of $f$:
\be
 \partial_{\theta} f = 
  \frac{\cos\theta}{\sin\theta} - \frac{b}{w} \, \rho^p  \cos^{p-1} \theta \sin\theta
   \;,\qquad \partial_{\rho} f= \frac{1}{\rho}  - \rho + \frac{b}{w} \,
    \rho^{p-1} \cos^p\theta + \frac{1}{w^2} \,\rho^{2p-1}  \; ,
\ee
and writes the saddle point equations as:
\be\label{eq:saddle1}
 \frac{b}{w} \, \rho^p   \cos^{p} \theta  = 
 \frac{\cos^2 \theta}{\sin^2\theta} \;,\qquad
 \frac{1}{\sin^2 \theta} - \rho^2 + \frac{1}{w^2} \rho^{2p}  =0 \Rightarrow \rho^2 = \frac{1}{\sin^2 \theta } T_p (  w^{-2} \sin^{2-2p} \theta )  \;  ,
\ee
with $T_p$ the Fuss--Catalan function.  In this form it is obvious that the saddle point equations always admit the solution $\theta_0 = \pi/2$, 
$\rho_0^2 = T_p(w^{-2})$ which reproduces the resolvent of the model with no spike. 
They admit at most one more solution $(\theta_1,\rho_1^2)$ with $\cos\theta_1\neq 0$. The dominant saddle point $\star$ is the one with the largest real part of:
\be
 f(\theta_i,\rho_i^2) = \frac{1}{2} \ln \bigg( T_p (w^{-2} \sin^{2-2p} \theta_i) \bigg) -\frac{p-1}{2p }  \;
 \frac{1}{\sin^2 \theta_i}T_p (w^{-2} \sin^{2-2p} \theta_i) 
 + \frac{1 - 2\sin^2 \theta_i  }{\sin^2 \theta_i} \;.
\ee 
The resolvent is given by the contribution of the leading saddle point $\star$, which is:
\be
 \omega(w) = \frac{1}{w} - p \frac{d}{dw} f(\theta_\star,\rho_\star) 
  =\frac{1}{w} \rho_{\star}^2
= \frac{1}{w \,\sin^2\theta_{\star}} 
T_p(w^{-2}\sin^{2-2p} \theta_{\star})  \;.
\ee

Irrespective of $b$, for $w\sim \infty$ only the saddle 
$(\theta_0, \rho_0^2 )$ stays at finite distance from the origin. We expect that for $b$ small enough this is the only viable saddle all the way up to $w\sim 0$. In this case the resolvent displays a cut\footnote{As $T_p(u)$ has a cut at $[u_c,\infty)$ with $u_c = (p-1)^{p-1}/p^p$,  $T_p(u_c) = p/(p-1)$. } on the real axis at $ [u_c^{-1/2},u_c^{1/2} ]$, identical to the model with no spike. Contrary to this, we expect that for $b$ larger than some threshold value $b_t$, when letting 
$w$ go to lower values the dominant saddle switches from $(\theta_0,\rho_0^2)$ to $(\theta_1,\rho^2_1)$. If this happens the singular locus of the resolvent changes. We interpret this as the detection threshold for the spike. 

We aim to identify the signal to noise ratio $b$ at which detection occurs. As we are interested in understanding the singular locus of $\omega(w)$, we restrict to $w = y \in \mathbb{R}_{+}$. 
The resolvent will always display a non removable singularity inherited from the end point of the cut of the Fuss Catalan function. Let us call $y_c$ this non removable singularity and let us denote $(\theta_c,\rho^2_c)$ the dominant saddle point at $y_c$.
We aim to find $\theta_c$ as a function of $b$:  we expect that for $b$ small $\theta_c =\pi/2$, while for $b$ large $\theta_c \neq \pi/2$.
In both cases $y_c$ respects:
\be
 y_c^2 \sin^{2p-2} \theta_c = u_c^{-1} 
 \Rightarrow \qquad
   \sin^{2}\theta_c=y_c^{ - \frac{2}{p-1}} \; \frac{p^{\frac{p}{p-1}}}{p-1}
  \;,\quad \rho^2_c = y_c^{\frac{2}{p-1}} u_c^{\frac{1}{p-1}}T_p(u_c)
 =y_c^{\frac{2}{p-1}} \;p^{ - \frac{1}{p-1}} \;.
\ee
If $\theta_c = \pi/2$ then $y_c = p^{p/2} /(p-1)^{(p-1)/2}$ and 
$\rho_c^2 = p/(p-1) $. If $\theta_c \neq\pi/2$ then, using the saddle point equations in Eq.~\eqref{eq:saddle1}, we get:
\be\label{eq:truc}
 \cos^{2p-4} \theta_c = \frac{y_c^2}{b^2 \rho_c^{2p} \sin^4\theta_c}
= \frac{1}{b^2} \;y_c^{\frac{2}{p-1}}
 \;(p-1)^2 p^{ - \frac{p}{p-1}} = 
  \left( 1 - y_c^{ - \frac{2}{p-1}}  \; \frac{p^{\frac{p}{p-1}}}{p-1} \right)^{p-2}
 \; .
\ee
We denote $ v =  y_c^{ - \frac{2}{(p-1)(p-2)}}
\;(p-1)^{ - \frac{2}{p-2}} p^{\frac{p}{(p-1)(p-2)}} $ and rewrite this as $h(v)=0$ with:
\be
h(v) = 1 - (p-1) v^{p-2} - b^{- \frac{2}{p-2}} \, v^{-1}  \;.
\ee

For $b$ small $h(v) = 0$ does not have real solutions. 
It follows that in this case the only viable saddle is 
$(\theta_c , \rho_c^2) = (\theta_0 , \rho_0^2)$. The detection threshold $b_t$ is the value at which Eq.~\eqref{eq:truc} develops real solutions, that is the lowest $b$ such that there exists $v_m$ with $h'(v_m) = h(v_m) = 0$. We 
get\footnote{Requiring
 $h'(v_m) =0$ fixes 
$
 v_m = (p-1)^{-\frac{1}{p-1} } \, (p-2)^{-\frac{1}{p-1}}
 \,  b^{-\frac{2}{(p-1)(p-2)}}
$
and imposing $h(v_m)=0$ yields $b_t$.}:
\be
 b_t^2 = \frac{(p-1)^p}{(p-2)^{p-2}} \; .
\ee
At $b= b_t$ we have $y_c = p^{p/2}$ which leads to 
$(\theta_c , \rho_c^2) = (\theta_1 , \rho_1^2)$
 with $\sin^{2}\theta_1 = \frac{1}{p-1}$ and $\rho_1^2 = p $. Finally, at $b=b_t$ and $y=p^{p/2}$ the saddle $(\theta_1 , \rho_1^2)$ is indeed dominant as: 
 \be
 \begin{split}
  f(\theta_0,\rho_0^2) & = \frac{1}{2} 
  \ln \bigg( T_p( p^{-p} ) \bigg)
   -\frac{p-1}{2p} T_p( p^{-p} ) -1 \crcr
   f(\theta_1,\rho_1^2) & =\frac{1}{2} \ln (\frac{p}{p-1})
    - \frac{p-1}{2p} (p-1) \frac{p}{p-1} + \left(1 - \frac{2}{p-1} \right) (p-1) 
    = \frac{1}{2} \ln (\frac{p}{p-1}) +\frac{p-5}{2} \; ,
 \end{split}
 \ee
and $T_p(u) \le T_p(u_c) = p/(p-1)$ for $u<u_c$, thus $f(\theta_1,\rho_1^2) >   f(\theta_0,\rho_0^2) $.

We conclude that, for $p\ge 3$, when dialing $b$ up from zero the largest non removable singularity of the resolvent jumps from $\sqrt{ p^{p}/(p-1)^{p-1}}$ to $p^{p/2}$ when $b$ reaches the threshold value $b_t$.

%%%%%%%%%%%%%%%%%%%%%%%%%%%%%%%%%%%%%%%%%%%%%%%%%%%%%%%%
%%%%%%%%%%%%%%%%%%%%%%%%%%%%%%%%%%%%%%%%%%%%%%%%%%%%%%%%
\section*{Acknowledgments}
%%%%%%%%%%%%%%%%%%%%%%%%%%%%%%%%%%%%%%%%%%%%%%%%%%%%%%%%
%%%%%%%%%%%%%%%%%%%%%%%%%%%%%%%%%%%%%%%%%%%%%%%%%%%%%%%%

The author would like to thank Manfred Salmhofer for pointing out references \cite{cartwright2013number} which inspired this project. The author is especially indebted to Dario Benedetti for many discussions and suggestions.

This work is supported by the European Research Council (ERC) under the European Union's Horizon 2020 research and innovation program (grant agreement No818066) and partly supported by Perimeter Institute for Theoretical Physics.

 \newpage
 
 \appendix

\section{The $\phi^p$ model in zero dimension}
\label{sec:toymodel}

In this section we present the scalar $\phi^p$ model defined by the partition function:
\be\label{eq:toyaction}
\begin{split}
Z(g) = \int_{\mathbb{R}}  [d\phi]   \; e^{- S(\phi) } \;,\qquad 
 S(\phi) = \frac{\phi^2}{2} - g^{\frac{p-2}{2}} \frac{\phi^p}{p} \; ,
\end{split}
\ee
where $g = |g| e^{\im \alpha}$ with $ 0 \le \alpha <2 \pi$ is a coupling constant and the measure $[d\phi] = (2\pi)^{-1/2} d\phi$ is normalized such that $Z(0)=1$. We are interested in the analytic structure of $Z(g)$ as a function of $g$ for $p\ge 3$. 
As a prerequisite, let us recall the Nevanlinna--Sokal theorem.

\begin{theorem}[Nevanlinna--Sokal]
A function $Z(g)$ is said to be Borel summable in $g$ along $\mathbb{R}_+$ if it is analytic in a disk tangent to the imaginary axis $D_R = {\rm Re}(g^{-1}) > R^{-1} $ and for $g\in D_R$ its Taylor expansion at the origin has a factorially bounded rest term:
\be
 Z(g) = \sum_{k\ge 0} a_k \, g^k   \;,
 \qquad \bigg|Z(g) - \sum_{k=1}^{n-1} a_k \, g^k\bigg| \le K \; n! \; A^n |g|^n\; , \;\;\forall g\in D_R \;,
\ee
with $K$ and $A$ some constants. If $Z(g)$ is Borel summable in $g$ along $\mathbb{R}_+$, then the series:
\be
 B(t) = \sum_{k\ge 0} \frac{a_k}{k!}  \,  t^k \;,
\ee
is absolutely convergent for $|t|<A^{-1}$ and admits an analytic continuation to the strip $\inf_{x\in \mathbb{R}_+}|t-x|<A^{-1} $ such that in the strip $ | B(t) | \le B  |\exp\{t/R\}|$. Furthermore, for any $g\in D_R$ the original function is given by the convergent integral: 
\be\label{eq:BRS}
 Z(g) = \int_0^{\infty} \frac{dt}{g} \; e^{-\frac{t}{g}} B(t) \;.
\ee
\end{theorem}

The function $B(t)$ is called the Borel transform of $Z(g)$, and $Z(g)$ is called the Borel sum of its perturbative series. The Borel transform is essentially an inverse Laplace transform, that is for $Z(g)$ analytic in $D_R$ and Borel summable along $\mathbb{R}_+$ we have:
\be
 B(t) = \frac{1}{2\pi \im} 
 \int_{ {\rm Re} (z^{-1} ) \le R^{-1}  } 
   \frac{ dz }{z}  \; e^{t/z} Z\left(z\right)  = 
    \frac{1}{2\pi \im} \int_{R^+-\im \infty}^{R^++\im\infty}  dz \;
     e^{zt} Z(z^{-1})
   \; ,
\ee
where these integrals converge for $t\in \mathbb{R}_+$.

As stated this theorem privileges the positive real axis. Tilting $g$ and $t$ in the complex plane leads to directional Borel summable functions and directional Borel transforms.

One can attempt to Borel sum any divergent series with factorial growth $a_n\sim n!$. The Borel transformed series $B(t) = \sum_k (k!)^{-1}a_k  \, t^k$
will have a finite radius of convergence and in simple cases one can analytically extend by hand $B(t)$ to some strips and compute the associated Borel sum $Z(g)$ using Eq.~\eqref{eq:BRS}. However, usually it is exceedingly difficult to find the analytic continuation of $B(t)$ by hand. What one can do in general is to show that the generating function $Z(g)$ of the series is Borel summable which implies that the analytic continuation of $B(t)$ to the strip exists. This requires to have a closed form expression for $Z(g)$ such that one can check the analyticity in the proper domain and the factorial bound on the rest term directly.

Going back to our partition function, we divide the plane of $g$ into angular sectors with aperture $\omega=\frac{2\pi}{p-2}$:
 \be
  S_q  = \bigg\{ g = |g|e^{\im \alpha}\in \mathbb{C} \; \bigg| \;\; 
  q \, \omega < \alpha < (q+1) \, \omega \bigg\} \;,\qquad q = 0,\dots p-3 \; ,
 \ee
separated by the roots of order $p-2$ of the unity $e^{\im q\omega}$.
Denoting $\eta =1$ for $p$ odd and $\eta=2$ for $p$ even, we observe that the partition function is periodic in the angular direction with period $\eta\omega$, that is $Z(e^{\im \eta\omega} g) = Z(g)$. The perturbative expansion of $Z(g)$ around $g=0$ is:
\be
 Z(g) = \sum_{n\ge 0} \frac{1}{n!} \frac{1}{p^n} 
  g^{n\frac{p-2}{2} } \int_{\mathbb{R}} [d\phi] \; e^{-\frac{\phi^2}{2}} \phi^{np}
   = \sum_{n\ge 0}^{np \text{ even}} \frac{(np)!}{n! (\frac{np}{2})! } \frac{1}{ ( 2^{\frac{p}{2}} p)^n }
  \;  g^{n\frac{p-2}{2}} \; ,
\ee
which is an integer series in $g$ with Borel transform:
\be\label{eq:Bseries}
 B(t) = \sum_{n\ge 0}^{np \text{ even}} \frac{(np)!}{n! (\frac{np}{2})! 
 ( n \frac{p-2}{2})!
 } \frac{1}{ ( 2^{\frac{p}{2}} p)^n }
  \;  t^{n\frac{p-2}{2}} \;. 
\ee

\subsection{Estimating the Borel transform}

The function $Z(g)$ can be reconstructed by directional Borel resummations. If $B(t)$ admits an analytic continuation in a strip around the ray
$t = e^{\im \alpha} |t|$ and in this strip it is bounded by 
$|B(t)| < B e^{|t|/R'}$, then the integral:
\be
 Z(g) = \int_{e^{\im \alpha} \mathbb{R}_+} \; \frac{dt}{g} \, e^{-t/g} 
 B( t ) \;,
\ee
is convergent for $g$ in a tilted disk $D_{R'}^{\alpha} = e^{\im \alpha} D_{R'}$ and defines a function $Z(g)$ which is Borel summable along the direction $\alpha$. However, as we will explain below, $Z(g)$ is not necessarily the Borel sum of the series for all $g\in D_{R'}^{\alpha}$. This is guaranteed only for $g$ along the original direction $g= |g| e^{\im \alpha}$.
Observe that $Z(g)$ reconstructed in this way depends on the directions along which $B(t)$ can be continued analytically:
if $B(t)$ has some singularity (pole or branch point) at $t_0$, then the directional Borel sum is discontinuous when the integration contour $e^{\im\alpha}\mathbb{R}_+$ passes the singularity.

The series $B(t)$ in Eq.~\eqref{eq:Bseries} is a series in $t^{\frac{p-2}{2}}$ for $p$ even and a series in $t^{p-2}$ for $p$ odd. As expected, it is periodic in the angular direction with period 
$\eta\omega$, $B(e^{\im \eta\omega} t) = B(t)$. Separating the term with $n=0$ and using Stirling's approximation
$n! \approx \sqrt{2\pi n} \; (n/e)^n$ we get:
\be
 B(t) \approx 1 + \,\frac{1}{\pi \sqrt{p-2}} \sum_{n\ge 1}^{np \text{ even}} 
  \frac{1}{n} \;\left[  \left( \frac{2p}{p-2} t \right)^{\frac{p-2}{2}}  \right]^n
   = 1 - \, \frac{\eta}{2\pi\sqrt{p-2}} \ln\left[ 1 - \left( \frac{2p}{p-2} t \right)^{\frac{p-2}{\eta}} \right] \;.
\ee

This function has $(p-2)/\eta$ cuts in the complex plane located at $t =    \rho   e^{\im  \eta \omega q } , \; \rho > \frac{p-2}{2p}$. For $p$ even ($\eta=2$), we get a cut every second sector, while for $p$ odd ($\eta=1$) we get a cut at the boundary of every sector. For $g=|g|e^{\im \alpha}\;, \alpha \neq \eta \omega$ the original partition function $Z(g)$ is then:
\be
 Z(g)  = \int_{e^{\im \alpha}\mathbb{R}_+} \frac{dt}{g}\,e^{-t/g}\, B(t) \approx 1 + \frac{\sqrt{p-2}}{2\pi} \int_{e^{\im \alpha} \mathbb{R}_+}
   du \; e^{- \frac{1}{g} \left( \frac{p-2}{2p}\right) u }
    \frac{  u^{\frac{p-2}{\eta} -1}  }{ 1 - u^{\frac{p-2}{\eta}} } \;,
\ee
where we integrated by parts. The discontinuities of $Z(g)$ at the cuts,
$Z ( |g|  e^{\im  \eta\omega^+  }) -  Z ( |g|  e^{\im \eta \omega^-} )$, which are all equal by periodicity, can be computed by deforming the contour to a (positively oriented) circle $C$ turning around the pole $e^{\im \eta \omega}$:
\be\label{eq:structdics}
\begin{split}
 {\rm Disc}_{ \eta \omega}(  |g| )  & = - \frac{\sqrt{p-2}}{2\pi} \int_{C} d u \;\frac{e^{ - \frac{1}{g} \left( \frac{p-2}{2p}\right) u } \; u^{ \frac{p-2}{\eta} -1 }}{ 
    (e^{\im \eta \omega} - w) \big[ e^{\im \eta \omega \left( \frac{p-2}{\eta} -1 \right) } + e^{\im \eta \omega \left( \frac{p-2}{\eta} -1 \right) } w + \dots + w^{\frac{p-2}{\eta}-1} \big]
  }\crcr
  & = \frac{ \im  \eta}{ \sqrt{p-2} }  \; e^{ -\frac{1}{|g|} \;\frac{p-2}{2p}} \;.
\end{split}
  \ee

\subsection{Borel summability} 
The drawback of the approach in the previous section is that, in order to study $Z(g)$, one needs to analytically continue $B(t)$ by hand. This is not possible in more complicated cases and we need a direct method to study the analyticity properties of $Z(g)$.

In each sector $S_q$ we define a sector partition function $Z_q(g)$ having the same perturbative expansion as $Z(g)$. We show that 
$Z_q(g)$ is Borel summable in the sector $S_q$, that is $Z_q(g)$ is analytic in $S_q$ extended by $\pm \pi/2$ and has a factorially bounded rest term.
Consequently $Z_q(g)$ is the Borel sum of the perturbative series in the sector $S_q$ hence it is the unique Borel summable function in $S_q$ with the right asymptotic series.

The analyticity domains of the Borel sums $Z_q(g)$ overlap, as 
$ (S_{q} \pm \pi/2 ) \cap S_{q+1} \neq 0$. But $Z_q(g)$ might be, and sometimes is, different from $Z_{q+1}(g)$ which naively contradicts the fact that the Borel sum of a series, if it exists, is unique. 
This is in fact not surprising: $Z_q(g)$ is the Borel sum in the sector $S_q$, and is the unique function \emph{Borel summable in $S_q$} with the right perturbative expansion around $0$. Now $Z_q(g)$ can be analytically continued outside $S_q$ into its neighbouring sector $S_{q+1}$,
but it ceases to be \emph{Borel summable} in $S_{q+1}$. In fact Borel summability at a point $g=|g|e^{\im \alpha}$ requires analyticity in a full disk 
$D^{\alpha}_R$. Although $Z_q(g)$ can be analytically continued to some $g\in S_{q+1}$, it can not be further continued to the corresponding disk $D^{\alpha}_R$. The Borel sum of the perturbative series in $S_{q+1}$ is in fact $Z_{q+1}(g)$ which is unique and can be different from $Z_q(g)$.

Finally, we compute the discontinuities of the $Z_q$s at the boundary of the sectors in terms of \emph{instantons}, that is non trivial solutions of the equations of motion of the action. 

Although $Z(g)$ is periodic and we could restrict to just the first $\eta$ sectors,  as it costs nothing to study an arbitrary sector, we will keep $q$ arbitrary below. 
 We promote the field $\phi$ to the complex plane:
\be 
\begin{split}
 Z(g) = \int_{C} [d\phi] \; e^{-S( \phi )}  \;,\qquad 
 S(\phi) = \frac{\phi^2}{2}  -   g^{\frac{p-2}{2}}\frac{\phi^p}{p} \;,
\end{split}
\ee
where the contour $C$ is a one dimensional contour from complex infinity to complex infinity. 

\paragraph{The sector partition functions.} 

We denote $\alpha_q = (q+1/2) \, \omega$ the bisectrix direction in the sector $S_q = \{ |g|e^{\im \alpha} | \, q\omega <\alpha <(q+1)\omega \}$. 
 Observe that for $p$ odd the roots sit opposite bisectrices, while for $p$ even the roots (respectively bisectrices) sit opposite each other\footnote{For $p=3$ we have $\omega =2\pi$ and only one sector $S_0=\{0 < \alpha < 2\pi\}$ with bisectrix $\alpha_0 = \pi$,
 while for $p=4$ we have $\omega = \pi$ and two sectors 
 $S_0 = \{0 < \alpha < \pi\}$ and $S_1 = \{ \pi <\alpha <2\pi\} $ with bisectrices  $\alpha_0 = \pi/2$ and $\alpha_1 = 3\pi/2$.}.
For $g\in S_q$ we define the sector partition function:
\be
    Z_q(g) =  \int_{e^{\im \theta_q} \mathbb{R}} \; [d\phi] \; 
  \exp\bigg\{ - \bigg( \frac{\phi^2}{2} - g^{\frac{p-2}{2}} \frac{\phi^p}{p}\bigg) \bigg\} \;,
\ee
where the domain of integration is the real axis, from $-\infty$ to $\infty$, tilted by $\theta_q$.
The tilting angle $\theta_q$ depends on the argument $\alpha$ of 
$g= |g|e^{\im \alpha}$ and is set to:
\be
 \theta_q = \frac{p-2}{2p} (\alpha_q -\alpha) \; \Rightarrow
 \qquad p\theta_q + \frac{p-2}{2}\alpha = \left( q+ \frac{1}{2}\right)\pi \;.  
\ee
A change of variables yields the integral formula:
\be
   Z_q(g)  = \int_{-\infty}^{\infty} e^{ \frac{\im}{2}  \frac{p-2}{p} 
  (\alpha_q - \alpha  )  }  [d\phi] \;
  \exp \bigg\{ - \left(\frac{ e^{\im \frac{p-2}{p} (\alpha_q -\alpha) } \phi^2 }{2}  -  \im (-1)^q |g|^{\frac{p-2}{2}} \frac{ \phi^p }{p}    \right) \bigg\} \;, 
\ee
which is absolutely convergent as long as
$ |\alpha_q -\alpha| < \pi/2 + \pi/(p-2) $, that is 
 $g$ belongs to the sector $S_q$ extended by $\pm \pi/2 $\footnote{
 In particular:
 \begin{itemize}
  \item[-] for $p=3$ we have $\theta_0 =  \frac{\pi -\alpha}{6} $ and: 
\be
 Z_0 (g)= \int e^{\im \frac{\pi-\alpha}{6}} [d\phi]
 \;   \exp\bigg\{ -  \bigg(  e^{\im \frac{\pi-\alpha}{3}}\frac{\phi^2}{2} - \im  \sqrt{|g|} \frac{\phi^3}{3}  \bigg ) \bigg\} \; , 
  \qquad \text{conv. for} \;\; -\frac{\pi}{2} < \alpha < 2\pi+\frac{\pi}{2} \;. \nonumber
\ee

\item[-] for $p=4$ we have $\theta_0 =  \frac{\pi}{8} - \frac{\alpha}{4} $, respectively $\theta_1 =  \frac{3\pi}{8} - \frac{\alpha}{4}$  and:
\be
\begin{split}\nonumber
 Z_0 (g) & = \int  e^{ \frac{\im}{2} \;\frac{1}{2} (\frac{\pi}{2} -\alpha) } 
 [d\phi ]\;   \exp\bigg\{ - \left(  e^{\im \;\frac{1}{2} (\frac{\pi}{2} -\alpha) }\frac{\phi^2}{2} -\im |g| \frac{\phi^4}{4}  \right) \bigg\} \; , \quad
  \text{conv. for} \;\; -\frac{\pi}{2} < \alpha < \pi +\frac{\pi}{2} \; ; \\
Z_1 (g)& = \int e^{ \frac{\im}{2} \;\frac{1}{2} (\frac{3\pi}{2} -\alpha) } [ d\phi ]
 \;   \exp\bigg\{ - \left( e^{\im \;\frac{1}{2} (\frac{3\pi}{2} -\alpha) }   \frac{\phi^2}{2} + \im |g| \frac{\phi^4}{4}  \right) \bigg\} \; ,
 \quad \text{conv. for} \;\; \frac{\pi}{2} < \alpha < 2\pi+\frac{\pi}{2} \;.
\end{split}
\ee
 \end{itemize}
}.

\paragraph{Analyticity.} The sector partition functions $Z_q(g)$ are analytic as long as they converge as they obey the Cauchy Riemann equations 
$\partial_{\bar g} Z_q(g) = 0$ with 
$\partial_{\bar g} = e^{\im \alpha} ( \partial_{|g|} + \frac{\im}{ |g| } \partial_{\alpha} ) $ in the domain of convergence:
\be
\begin{split}
 e^{-\im \alpha}\partial_{\bar g} Z_q(g) & = \int_{-\infty}^{\infty}
e^{  \frac{\im }{2} \frac{p-2}{p} (\alpha_q - \alpha ) } [d\phi] \; 
 \exp\bigg\{ - \bigg( e^{  \im 
  \frac{p-2}{p} (\alpha_q - \alpha) }\frac{\phi^2}{2} - \im (-1)^q  
    |g|^{\frac{p-2}{2}} \frac{\phi^p}{p} \bigg)\bigg\} \crcr
    & \qquad \qquad \bigg[  \frac{ \frac{p-2}{2} } {|g|}
     \im (-1)^q   |g|^{\frac{p-2}{2} }\frac{\phi^p}{p} 
    + \frac{ \im  }{|g|}
     \bigg\{ - \im \frac{p-2}{2p} 
       + \im \frac{p-2}{p}   e^{\im \frac{p-2}{p} (\alpha_q -\alpha) } \frac{\phi^2}{2}
     \bigg\}
    \bigg] \crcr
& =  \frac{p-2}{2p|g|} \int_{-\infty}^{\infty} [d\phi] \frac{\delta}{\delta \phi} 
    \bigg(\phi e^{-S(\phi)} \bigg)=0 \; .
\end{split}
\ee

\begin{figure}[ht]
\begin{center}
\psfrag{S}{$S_q$}
\psfrag{q}{$q\omega$}
\psfrag{q1}{$(q+1)\omega$}
\psfrag{a}{$\alpha_q$}
\includegraphics[width=0.5\textwidth]{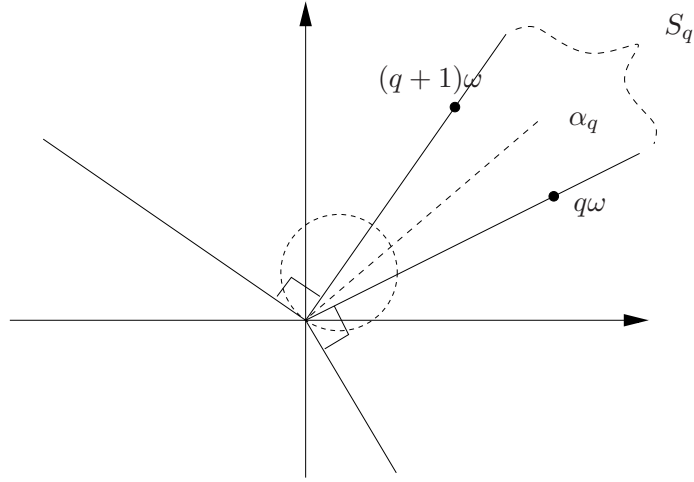}
 \caption{The plane of the coupling constant $g$. The sector $S_q$, its bisectrix and the domain of analyticity of $Z_q(g)$. Observe that a disk $D^{\psi}_R$ can fit in the domain of analyticity only for $ q\omega < \psi <(q+1) \omega$, that is $Z_q(g)$ is the Borel sum of the series only in $S_q$. } \label{fig:finalpicture}
 \end{center}
\end{figure}

\paragraph{Perturbative series and Taylor rest.} 
As expected, the sector partition functions $Z_q(g)$ have the same perturbative series:
\be 
Z_q(g) =  \int_{e^{\im \theta_q} \mathbb{R}} \; [d\phi]  \; 
  \exp\bigg\{ - \bigg( \frac{\phi^2}{2} - g^{\frac{p-2}{2}}  \frac{\phi^p}{p}\bigg) \bigg\} =   \sum_{n\ge 0}^{np \text{ even}}  \frac{(np)!}{ n! (np/2)!  }  \; \frac{1}{ (2^{p/2} p)^n   } \; g^{n\frac{p-2}{2}} \;,
\ee
obtained by Taylor expanding the interaction and computing the Gaussian integrals. In order to prove directional Borel summability we must check that the Taylor rest terms are factorially bounded. Using a Taylor expansion with integral rest:
\be
 f(1) = \sum_{q=0}^{n-1} \frac{1}{q!} f^{( q ) }(0)
 +  \frac{1}{(n-1)! } \int_0^1 du \;  (1-u)^{n-1} f^{(n) }(u) \;,
\ee
and putting the interpolation parameter on the interaction term we find that the Taylor rest term $R_q^{(n)}(g) = Z_q(g) - \sum_{k}^{(n-1)\frac{p-2}{2}} a_kg^k$  is :
\be 
 R_q^{(n)}(g)   = 
 \frac{1}{(n-1)! } \int_0^1 du \;  (1-u)^{n-1}  \int_{e^{\im \theta_q} \mathbb{R}} \;[d\phi]   \; \exp\bigg\{ - \bigg( \frac{\phi^2}{2} -  u g^{\frac{p-2}{2}} \frac{\phi^p}{p}\bigg) \bigg\} \left( g^{\frac{p-2}{2}}\frac{\phi^p}{p} \right)^n \;,
\ee
where either $p$ is even (and $n$ is unconstrained) or $n$ is only even. Taking absolute values we note that along the direction of integration 
$\exp( u g^{\frac{p-2}{2}}\frac{\phi^p}{p } )$ is uniformly bounded by $1$. Integrating out $u$ we get a bound on the rest term:
\be
\begin{split}
\big|R_q^{(n)}(g) \big| & \le \frac{1}{n!} \;\frac{|g|^{\frac{p-2}{2} n }}{p^n} 
 \int_{-\infty}^{\infty}   [d\phi]  \;
  \exp\bigg\{ -\cos\bigg[\frac{p-2}{p} (\alpha_q-\alpha) \bigg] \frac{\phi^2}{2}\bigg\}   \phi^{np} \crcr
  & \qquad =\frac{1}{n!} \;\frac{ |g|^{n\frac{p-2}{2} } }{p^n} \; \frac{(np)!}{ 2^{\frac{np}{2} } (np/2)!  }  \;  \frac{1}{ 
  \bigg( \cos\bigg[ \frac{p-2}{p} (\alpha_q-\alpha) \bigg] \bigg)^{np+1/2}  }  \;.
\end{split}
\ee
This is factorially bounded for $g = |g| e^{\im \alpha}$ as long as
$ \left| \frac{p-2}{p} (\alpha_q-\alpha) \right| <\frac{\pi}{2}$, that is 
$g$ belongs to the sector $S_q$ extended by $\pm \pi/2$.
It follows that $Z_q(g)$ is Borel summable in $g$ in the sector $S_q$ (more precisely along all the rays in the sector $S_q$) and can be reconstructed in $S_q$ extended by $\pm\pi/2$ by the  convergent Laplace transform: 
\be
 Z_q(g) = \int_{e^{\im \alpha} \mathbb{R}_+} \frac{dt}{g} \;
  e^{-t/g} B(t) \;, \qquad g= |g| \, e^{\im \alpha} \;, \;\;
 q\omega  - \frac{\pi}{2}  <\alpha < 
 (q+1) \omega +\frac{\pi}{2} \;.
\ee
The final picture is presented in Fig.~\ref{fig:finalpicture}.

\paragraph{Discontinuities of the Borel sum.} 
At the interface of two neighbouring sectors the Borel sum of the series changes from $Z_{q-1}(g)$ to $Z_{q}(g)$. This change can either be continuous, that is the two sector partition function agree at $q\omega$, or not. We therefore have $p-2$ possible discontinuities:
 \be
  Z_q ( |g|  e^{\im q \omega^+  }) - 
    Z_{q-1} ( |g|  e^{\im q \omega^-} ) \; ,\qquad q=0,\dots p-3 \;,
 \ee
where $Z_{-1} \equiv Z_{p-3}$ and $0^-\equiv 2\pi^-$. 
As $  \theta_q = \frac{p-2}{2p} (\alpha_q -\alpha) $, when the argument of $g$ approaches the boundaries of a sector we get:
\be
 \theta_q\Big{|}_{\alpha \searrow q \omega} = \frac{\pi}{2p}
 \;,\qquad  \theta_{q-1}\Big{|}_{\alpha \nearrow q \omega}  = -\frac{\pi}{2p} \;,
\ee
and the discontinuity writes:  
\be
  Z_q ( |g|  e^{\im q \omega^+  }) - 
    Z_{q-1} ( |g|  e^{\im q \omega^-} )
  = \int_{(e^{\im\frac{\pi}{2p}} - e^{-\im\frac{\pi}{2p}} ) \mathbb{R}} d\phi \;
  \exp \bigg\{- \bigg( \frac{\phi^2}{2} - \left( |g| e^{\im \omega q} \right) ^{\frac{p-2}{2} } 
  \frac{\phi^p}{p} \bigg)\bigg\}  \;.
\ee
We reverse the orientation on $ e^{ -\im \frac{\pi}{2p}} \mathbb{R} $ in order to obtain a sum. We will compute the discontinuity by a saddle point approximation. 

The \emph{instantons} are the non trivial saddle points of the action  $ S(\phi) = \phi^2/2 - g^{( p-2)/2} \phi^p/p  $, that is the non trivial solutions of the equations of motion
$\partial_\phi  S (z)  = \phi - g^{(p-2)/2} \phi^{p-1} =0$. In our case the instantons are proportional to the roots of order $p-2$ of unity and for $g$ at the cut $q$ we obtain:
\be
 \phi_r = |g|^{-1/2} \;e^{\im \omega \left( r - \frac{q}{2} \right)} \;.
\ee

We cut and join the integration contour in $\phi$ respecting the orientations to obtain two integration contours (see Fig.~\ref{fig:finalpicture1} below):
\be
 C = e^{ - \im  \frac{\pi}{2p} } \infty  -  0 - e^{\im \frac{\pi}{2p}} \infty \;, \qquad 
\tilde C  = 
 e^{\im  \frac{\pi}{2p}} (- \infty) - 0 - e^{ - \im \frac{\pi}{2p}} (-\infty) \; ,
\ee
which we attempt to push towards complex infinity. If no instanton $\phi_r$ is traversed in the process the contour does not contribute. If an instanton $\phi_r$ is traversed, its contribution must be taken into account. We thus need to count how many instantons (if any) are traversed by the contours $C$ and $\tilde C$, that is how many $\phi_r$s sit in an angular sector of aperture $\pi/p$ centered on the real axis. Formally, we search for integers $r$ solutions of the equations:
\be
   -\frac{\pi}{2p} < \omega \left( r - \frac{q}{2} \right)   <
     \frac{\pi}{2p} \;, \qquad 
      \pi  -\frac{\pi}{2p} < \omega \left( r - \frac{q}{2} \right)   <
     \pi + \frac{\pi}{2p}  \;.
\ee
As $\omega ( r-q/2 )$ is a half integer multiple of $\omega$ and  
$ \pi /  p <  \omega$, the only possible solutions of these equations are:
\be
 \omega \left( r - \frac{q}{2} \right)  =0 \;,\qquad 
  \omega \left( r - \frac{q}{2} \right)  = \pi =\frac{p-2}{2} \omega  \;,
\ee
with $r$ integer, $0\le r \le p-3$. In particular this implies that $\phi_r$ contributes if and only if it is \emph{real}.
We have the following cases:
\begin{itemize}
 \item $C$ traverses $\phi_r$ with $r=q/2$ for $q$ even. 
 \item $\tilde C$ traverses $\phi_r$ with $r = \frac{p+q-2}{2}$ for 
 $p+q$ even.
 \end{itemize}
In order to compute the contribution of an instanton, we denote:
\be
I = \int_{C} [d\phi] \; \exp \bigg\{-  \bigg( \frac{\phi^2}{2} - g^{\frac{p-2}{2}} \frac{\phi^p}{p} \bigg)\bigg\} \; , \qquad
 \tilde I = \int_{\tilde C} [d\phi] \; \exp \bigg\{-  \bigg( \frac{\phi^2}{2} - 
 g^{\frac{p-2}{2}}  \frac{\phi^p}{p} \bigg)\bigg\}  \; .
\ee
A second order Taylor expansion around $\phi_r$ yields 
$
 S(\phi_r +  x) = \frac{ p-2}{2p} \phi_r^2   - (p-2)   x^2  + O(x^3)
$
therefore, as $C$ can only trap $z_r$ with $r=q/2$ for even $q$ while $\tilde C$ can only trap $\phi_r$ with $r = (p+q-2)/2$ for even 
$p+q-2$ we get:
\be
 I = \tilde I =  e^{-\frac{1}{|g|} \,\frac{p-2}{2p}} 
  \int_{-\im \infty}^{\im \infty} dx \; e^{\frac{p-2}{2} x^2 }
 = \frac{ \im }{ \sqrt{p-2} } \; e^{-\frac{1}{|g|} \,\frac{p-2}{2p}} 
 = \frac{1}{\sqrt{S''(\phi_r)}} e^{-S(\phi_r)}\; ,
\ee
where we took into account that both $C$ and $\tilde C$ traverse the saddles parallel to the imaginary axis from $-\im \infty$ to $\im \infty$. This reproduces exactly the discontinuities computed in Eq.~\eqref{eq:structdics}. The final picture is depicted in Fig.~\ref{fig:finalpicture1}\footnote{
For $p=3$ we only have one candidate discontinuity along the positive real axis corresponding to $q=0,p=3$. The instanton $\phi_0$ contributes and it brings $
 Z_0 ( |g|  e^{\im  0^+  }) - 
    Z_{0} ( |g|  e^{\im  2\pi^-} ) = \im \;e^{-\frac{1}{6|g|}}
$. For $p=4$ we have two candidate discontinuities, one along the positive real axis ($q=0,p=4$) and one along the negative real axis ($q=1,p=4$). For the first one both instantons $\phi_0$ and $\phi_1$ contribute
$ Z_0 ( |g|  e^{\im  0^+  }) - 
    Z_{1} ( |g|  e^{\im 2\pi^-} )  =  2 \frac{\im}{\sqrt{2}} \;e^{-\frac{1}{4|g|}} $, while for the second one no instanton contributes 
    $ Z_1 ( |g|  e^{\im \pi^+  }) - 
    Z_{0} ( |g|  e^{\im \pi^-} ) = 0 $.
}

\begin{figure}[ht]
\begin{center} 
\psfrag{z}{$z_{q/2}$} 
\psfrag{t}{$\pi/2p<\omega/2$} 
\includegraphics[width=0.5\textwidth]{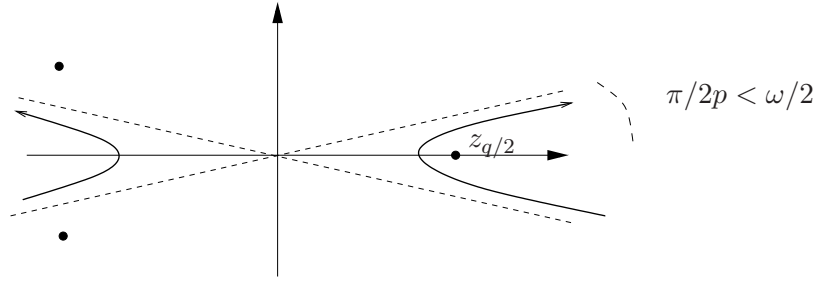}
 \caption{The plane of the field $\phi$. The case when the contour $C$ traps the saddle $\phi_{q/2}$ and the contour $\tilde C$ does not trap any saddle (that is $q$ even and $p$ odd).} \label{fig:finalpicture1}
 \end{center}
\end{figure}

\subsection{Rescaled coupling}

We are interested in the analytic structure of the partition function in terms of a rescaled coupling: 
\be
{\cal Z}(w) = Z(w^{-\frac{2}{p-2}}) = \int_{\mathbb{R}} [d\phi] \; e^{-{\cal S}(\phi)} \; , \qquad
  {\cal S}(\phi)  = \frac{\phi^2}{2}  - \frac{1}{w} \frac{\phi^p}{p} \; .
\ee
This can be deduced from the previous discussion in two steps. First  change variables to $w' = g^{\frac{p-2}{2}}$. This blows up the first two sectors (with aperture 
$2\omega = 2\frac{2\pi}{p-2} $) to the whole complex plane. In terms of $w'$ the partition function can have at most two cuts, one along the negative real axis and one along the positive real axis. Both cuts run from $0$ to complex infinity. The (possible) discontinuities at the cuts are the same ones we already computed, as $Z(g)$ is periodic with period $2\omega$. Second we pass to $w=1/w'$ which has the effect of reversing the cuts. 

Overall we conclude that ${\cal Z}(w)$ is analytic outside the real axis. It is convenient to parametrize $w = |w| e^{\im \psi}, \, -\pi <\psi<\pi$. We have the convergent integral representations:
\be
\begin{split}
0 <\psi<  \pi \;\qquad & {\cal Z}_+(w)= 
 \int_{e^{\im \theta_+}\mathbb{R}} [d\phi] \;\exp \bigg\{ - \left( \frac{\phi^2}{2} - \frac{1}{w} \, \frac{\phi^p}{p}\right) \bigg\} \;, 
  \qquad \theta_+ = \frac{1}{p} \left( \psi - \frac{\pi}{2}\right) \;,
 \\
-\pi < \psi< 0  \;\qquad & {\cal Z}_-(w)=
\int_{e^{\im \theta_-}\mathbb{R}} [d\phi] \;\exp \bigg\{ - \left( \frac{\phi^2}{2} - \frac{1}{w} \, \frac{\phi^p}{p}\right) \bigg\} \; ,
 \qquad \theta_- = \frac{1}{p} \left( \psi + \frac{\pi}{2}\right) \; .
\end{split}
\ee
${\cal Z}_+$ can be analytically continued for $|\psi-\pi/2| < p\pi/4$, while 
${\cal Z}_-$ can be analytically continued for $|\psi + \pi/2| < p\pi/4$. ${\cal Z}_{\pm}$ are the Borel-Leroy sums of order $(p-2)/2$ of their perturbative series. We have the integral representation:
\be
 {\cal Z}_{\pm}(|w|e^{\pm \im \psi}) 
  = \frac{2}{p-2} \int_{e^{-\im \psi}\mathbb{R}_+} \left(   w^{\frac{2}{p-2}} u^{\frac{2}{p-2}-1} du \right)   e^{  - (wu)^{\frac{2 }{p-2} } }  B(u^{2/(p-2)}) \; ,
\ee
with $B$ the Borel transform of the original partition function $Z(g)$.
As a function of $u$, $B(u^{2/(p-2)})$ is analytic in $\mathbb{C} \setminus \{ (-\infty, u_m) \cup[u_M,\infty) \}$ with $u_m<0<u_M$. Its singularities are either poles or branch points, but always lead to cuts for ${\cal Z}(w)$.
Thus ${\cal Z}(w)$ can be discontinuous only at the real axis. The discontinuity at a cut is (for $y\in \mathbb{R}$):
\be\label{eq:discinst}
     {\cal Z}_{+}( y ) -  {\cal Z}_{-}( y  ) 
      =\sum_{\nu} \frac{1}{\sqrt{ {\cal S}''(\phi_{\nu} ) }} e^{-{\cal S}(\phi_{\nu} ) } \;,
\ee
where $\phi_{\nu}$ are the real instantons, that is the solutions of the equation of motion ${\cal S}'(\phi_{\nu})=0$, which become real for $w = y $. While we derived this formula for the $\phi^p$ model, it is in fact general. The only ambiguity here is the determination of the square root in Eq.~\eqref{eq:discinst} which is fixed by checking carefully the steepest descent paths trough the instantons\footnote{
In the $\phi^p$ model, for $p$ even ${\cal Z}(w)  $ has a cut 
at $y\ge 0$ with discontinuity
${\cal Z}_+(  y ) -    {\cal Z}_- ( y ) 
=2 \im (p-2)^{-1/2} \exp\{ - \frac{p-2}{2p} y^{\frac{2}{p-2}} \} $ 
while for $p$ odd it has two cuts $  {\cal Z}_- ( -|y| ) -  {\cal Z}_+( - |y| )  =     {\cal Z}_+ ( |y| ) -  {\cal Z}_-( |y|  )
  = \im (p-2)^{-1/2} \exp\{ - \frac{p-2}{2p} |y|^{\frac{2}{p-2}} \}   
     $.}.

\newpage

%----- Bibliography ----------------------
 \bibliographystyle{JHEP-3}
 
  \bibliography{/home/razvan/Desktop/lucru/Ongoing/Refs/Refs.bib}
%---------------------------------------------
%------------------------------------------------------------------------------

%------------------------------------------------------------------------------
\end{document}